\documentstyle[prl,multicol,aps,epsfig]{revtex}
\newcommand \beq{\begin{eqnarray}}
\newcommand \eeq{\end{eqnarray}}
\def\simge{\mathrel{%
       \rlap{\raise 0.511ex \hbox{$>$}}{\lower 0.511ex \hbox{$\sim$}}}}
\def\simle{\mathrel{
       \rlap{\raise 0.511ex \hbox{$<$}}{\lower 0.511ex \hbox{$\sim$}}}}

\begin{document}

\title{Tkachenko modes of vortex lattices in rapidly rotating
   Bose-Einstein condensates}
\author{Gordon Baym}
\address{Department of Physics, University of Illinois at
Urbana-Champaign, 1110 West Green Street, Urbana, Illinois 61801}
\date{\today}
\maketitle

\begin{abstract}

    We calculate the in-plane modes of the vortex lattice in a rotating Bose
condensate from the Thomas-Fermi to the mean-field quantum Hall regimes.  The
Tkachenko mode frequency goes from linear in the wavevector, $k$, for lattice
rotational velocities, $\Omega$, much smaller than the lowest sound wave
frequency in a finite system, to quadratic in $k$ in the opposite limit.  The
system also supports an inertial mode of frequency $\ge 2\Omega$.  The
calculated frequencies are in good agreement with recent observations of
Tkachenko modes at JILA, and provide evidence for the decrease in the shear
modulus of the vortex lattice at rapid rotation.

\end{abstract}

\begin{multicols}{2}
\narrowtext

    The collective modes of the vortex lattice in a rotating superfluid have
been of considerable interest since the 1960's.  In a classic series of papers
Tkachenko \cite{Tkachenko} showed that the lattice supports an elliptically
polarized oscillatory mode, with the semi-major axis of the ellipse orthogonal
to the direction of propagation.  Such modes were observed in superfluid
helium in 1982 \cite{andereck}.  Reference~\cite{BC} reformulated the
hydrodynamics of rotating superfluids to take into account the elasticity of
the vortex lattice (including the normal fluid, dissipation, and line bending
-- Kelvin -- oscillations of the vortex lines in three dimensions) and thus
describe the Tkachenko modes; effects of the oscillations of the vortex lines
at finite temperature on the long ranged phase correlations of the superfluid
were discussed in \cite{lattice}.  The focus of these papers was on superfluid
helium, where rotational speeds are always much smaller than characteristic
phonon frequencies, of order 1K/$\hbar$.

    Atomic Bose condensates, on the other hand, allow one to study superfluids
over a large range of rotational speeds
\cite{JILA,Madison,VortexLatticeBEC,HaljanCornell,MITimprint}, from the
Thomas-Fermi regime where lattice rotational frequencies, $\Omega$, are small
compared with phonon frequencies -- of order the trapping frequency -- to the
quantum Hall limit of rapid rotations \cite{Cooper,Viefers,Jolicoeur,Ho},
where a condensate in a harmonic trap flattens to a very weakly interacting
effectively two dimensional system, and rotational speeds can well exceed
phonon frequencies.  One may identify three distinct physical regimes:  first,
the ``stiff" Thomas-Fermi regime, where $\Omega$ is small compared with the
lowest compressional frequencies, $sk_0$, where $s$ is the sound velocity,
$k_0\sim 1/R$ is the lowest wavenumber in the finite geometry, and $R$ is the
size of the system transverse to the rotation axis.  In this regime the system
responds to rotation effectively as an incompressible fluid.  At faster
rotation, when $\Omega \gg sk_0$, but $\Omega \ll ms^2$, where $m$ is the
atomic mass, the system is in the ``soft" Thomas-Fermi regime, where
compression of the superfluid becomes important in the response of the
lattice.  Finally, when $\Omega \gg ms^2$, the system enters the ``mean field"
quantum Hall regime, in which the condensation is only in lowest Landau
orbits.  Eventually the vortex lattice melts \cite{RS,SHM}, and the system
enters a strongly correlated regime \cite{Cooper,Viefers,Jolicoeur}.

    Observations of Tkachenko modes in Bose-condensed $^{87}$Rb were reported
recently by Coddington et al.~\cite{jilatk} at rotation speeds up to 0.975 of
the transverse trapping frequency, $\omega_\rho$; in this regime, $\Omega$
increases past $sk_0$, and although not yet in the quantum Hall regime, the
experiments show effects on the modes frequencies of very rapid rotation.  The
modes in rotating atomic condensates have been the subject of several
theoretical investigations, including determination of the Tkachenko modes at
slow rotation taking full account of the finite geometry ~\cite{crescimanno},
determination of the fundamental modes in the absence of effects of the
elastic energy of the lattice \cite{cozzini1}, and studies in the quantum Hall
regime \cite{SHM,cozzini2,cazalilla,qhmodes}.  In this Letter we derive the
modes of the vortex lattice at general rotation speeds.  The present analysis
is restricted to linearized motion in two dimensions transverse to the
rotation axis, and neglects the normal fluid \cite{drew}.  Our starting point
is the conservation laws and superfluid acceleration equation governing the
system, including the full elasticity of the lattice.  Taking into account the
decrease of the shear modulus of the vortex lattice with increasing $\Omega$,
one can fully understand the measured Tkachenko frequencies.

    We consider a rotating gas of bosons described by a repulsive short range
repulsive interaction, $U(r)= g\delta^3(r)$, with $g = 4\pi a_s/m$, where
$a_s$ is the s-wave scattering length, in units in which $\hbar=1$.  The
angular momentum of the superfluid is carried in vortices, which form a
triangular lattice rotating as a solid body at angular velocity $\Omega$.  We
work in the frame corotating with the vortex lattice. and denote the
deviations of the vortices from their home positions by the continuum
displacement field, $\epsilon(r,t)$.  In linear order in the vortex
displacements, the long wavelength superfluid velocity, $v(r,t)$, can be
written, following \cite{lattice}, in terms of the long wavelength vortex
lattice displacement field, $\epsilon(r,t)$, and the phase $\Phi(r,t)$ of the
order parameter, as
\beq
   v + 2\Omega\times\epsilon = \nabla \Phi/m.
 \label{vphase}
\eeq
This equation follows from the fact that its curl,
\beq
  \nabla\times v = -2\Omega \nabla\cdot\epsilon,
\label{curlv}
\eeq
is the conservation law relating the change in vorticity to a compression
of the vortex lattice, while its longitudinal part is trivially the gradient
of a scalar \cite{curl}.  Equation (\ref{curlv}) constrains the number of
degrees of freedom in two dimensions from five ($n$, $v$, $\epsilon$)
to four.  The time derivative of Eq.~(\ref{vphase}) is the superfluid
acceleration equation,
\beq
  m\left({{\partial }\over{\partial t}} + 2\Omega \,\times
    \dot{\epsilon}\right) = - \nabla (\mu - V_{\rm eff}),
  \label{supaccel}
\eeq
where $\mu$ is the chemical potential, and, for a harmonic confining trap
of frequency $\omega_\rho$ in two dimensions,
\beq
 V_{\rm eff} = \frac m2 (\omega_\rho^2-\Omega^2) r^2.
\eeq
In the frame corotating with the vortex lattice, the chemical potential $\mu$
is related to the phase by
\begin{eqnarray}
\mu(r,t) - V_{\rm eff} = -\frac{1}{m}\frac{\partial \Phi(r,t)}{\partial t}.
  \label{mu-phi}
\end{eqnarray}

     The local elastic energy density of a triangular lattice in two
dimensions has the form \cite{BC},
\beq
 {\cal E}(r) = 2C_1 (\nabla\cdot\epsilon)^2
        +C_2\left[\left(\frac{\partial \epsilon_x}{\partial x}
          -\frac{\partial\epsilon_y}{\partial y}\right)^2
          \right. \nonumber \\ \left.
    + \left(\frac{\partial \epsilon_x}{\partial y}  +\frac{\partial
     \epsilon_y}{\partial x}\right)^2\right],
 \label{elastic}
\eeq
where $C_1$ is the compressional modulus, and $C_2$ the shear modulus of
the vortex lattice.  In an incompressible fluid, $C_2 = n\Omega/8 = -C_1$.
The shear modulus $C_2$ in fact decreases with increasing $\Omega$, from
$\Omega n/8$ in the incompressible limit $\Omega$, eventually reaching, in the
mean field quantum Hall limit, the value \cite{SHM,qhmodes}, $C _2 \simeq
(81/80\pi^4)ms^2 n$.  In this limit $C_1 = 0$.  The falloff for small
$\Omega/ms^2$ has the form,
\beq
    C_2 \simeq \frac{\Omega n}{8}\left(1-\gamma
   \frac{\Omega}{ms^2}+\dots\right);
\label{C2drop}
\eeq
a first estimate \cite{qhmodes} is $\gamma \sim 4$.

    The dynamics is specified by the superfluid acceleration equation
(\ref{supaccel}); the continuity equation,
\beq
  \frac{\partial n(r,t)}{\partial t} + \nabla \cdot j(r,t) = 0,
  \label{contin}
\eeq
where $n$ is the (smoothed) density, and $j=nv$ is the particle current; and
conservation of momentum:
\beq
   m\left(\frac{\partial j}{\partial t} +  2\Omega\times j\right) +\nabla P
        + n\nabla  V_{\rm eff} = -\sigma,
  \label{momcons}
\eeq
where $P$ is the pressure; at zero temperature, $\nabla P = n\nabla \mu$,
while in equilibrium, $\nabla P +n\nabla V_{\rm eff} = 0$.  The elastic
stress, $\sigma$, is given in terms of the total elastic energy, $E_{\rm el} =
\int d^2r {\cal E}(r)$, by
\beq
 \sigma(r,t) = \frac{\delta E_{\rm el}}{\delta \epsilon}
               = -4C_1\nabla (\nabla\cdot\epsilon) -2C_2\nabla^2 \epsilon.
\eeq

    Equations~(\ref{momcons}) and (\ref{supaccel}), with $\nabla P =
n\nabla\mu$, imply $2\Omega\times({\dot \epsilon}-v) = \sigma/mn$.  The curl
of this equation becomes
\beq
   \nabla\cdot ({\dot\epsilon} - v)
   = \frac{\nabla\times\sigma }{2\Omega nm}
   = \frac{C_2}{\Omega nm}\nabla^2(\nabla\times\epsilon),
 \label{diveps}
\eeq
while its divergence, together with (\ref{curlv}), yields,
\beq
  \nabla\times {\dot \epsilon} +
         2\Omega  \nabla \cdot{\epsilon}
             = -\frac{\nabla\cdot\sigma}{2\Omega nm}
    =\frac{C_2+2C_1}{\Omega nm}\nabla^2(\nabla\cdot\epsilon).
 \label{xeps}
\eeq
The structure of the modes follows directly from the linearized set of
equations:  (\ref{curlv}), (\ref{diveps}), (\ref{xeps}), and the divergence of
(\ref{supaccel}).  The density oscillations are governed by
\beq
\left(-\frac{\partial^2}{\partial t^2} + s^2\nabla^2\right) n =
             2n\Omega\nabla\times{\dot \epsilon},
 \label{densosc}
\eeq
where the sound speed $s$ is given by $ms^2=\partial P/\partial n$
\cite{sound}.  In the absence of coupling of density oscillations to
compressions of the vortex lattice, this equation is simply that of long
wavelength phonons in the condensate.  Using Eq.~(\ref{diveps}) to eliminate
$\nabla\times\epsilon$, and ignoring the term of order $\nabla^4$, we
find
\beq
  \left(-\frac{\partial^2}{\partial t^2} + \frac{2C_2}{mn}\nabla^2\right)
     \nabla\cdot\epsilon = \frac{1}{n}\frac{\partial^2}{\partial
        t^2} n.
\label{epsosc}
\eeq
In the absence of coupling of density oscillations to compression of the
vortex lattice, this equation would describe a free Tkachenko mode of the
vortex lattice \cite{Tkachenko,BC,lattice} of wavevector $k$ and frequency,
$\omega_T = (2C_2/mn)^{1/2}k$.

    The coupled equations (\ref{densosc}) and (\ref{epsosc}) yield the full
spectrum of long wavelength modes.  The frequencies, for given $k$, are
solutions of the secular equation,
\beq
  D(k,\omega) \equiv &&\omega^4 - \omega^2\left[4\Omega^2 + \left(s^2
   +\frac{4}{nm}(C_1+C_2)\right)k^2\right] \nonumber\\
 &&  + \frac{2s^2C_2}{nm}k^4 =
   (\omega^2 -\omega_I^2)(\omega^2 -\omega_T^2) =0.
 \label{D}
\eeq
For $2s^2C_2k^4/nm \ll (4\Omega^2 + (s^2 +4(C_1+C_2)/nm)k^2)^2$, as is
generally the case, the mode frequencies are given by \beq
  \omega_I^2 = 4\Omega^2 + \left(s^2+\frac{4(C_1+C_2)}{nm}\right)k^2,
 \label{inertial}
\eeq
and
\beq
  \omega_{T}^2 = \frac{2C_2}{nm} \frac{s^2k^4}{(4\Omega^2 +
   (s^2+4(C_1+C_2)/nm)k^2)}.
  \label{tk}
\eeq
The first mode is the standard inertial mode of a rotating fluid; for
$\Omega \ll s^2k^2$ it is a sound wave, while for $\Omega \gg s^2k^2$, the
mode frequencies begin essentially at $2\Omega$.  The second mode is the
observed elliptically polarized Tkachenko mode.  See Fig.~1.  As
Eq.~(\ref{xeps}) implies, the longitudinally polarized component is $\pi/2$
out of phase with the transversely polarized component; of order
$\omega_T/2\Omega$ smaller for the Tkachenko mode and of similar size for the
inertial mode \cite{3d}.  In the stiff limit, the Tkachenko frequency,
$\omega_T=(\Omega/4m)^{1/2}k$, is linear in $k$.  By contrast, in the very
soft limit, the mode frequency, $\omega_T$ is quadratic in $k$ at long
wavelengths
\beq
  \omega_T = \left(\frac{C_2}{2nm}\right)^{1/2} \frac{sk^2}{\Omega};
\eeq
in the quantum Hall regime $\omega_T \simeq
(9/4\pi^2\sqrt{10})(s^2k^2/\Omega)$ \cite{SHM}.

\begin{figure}
\begin{center}
\epsfig{file=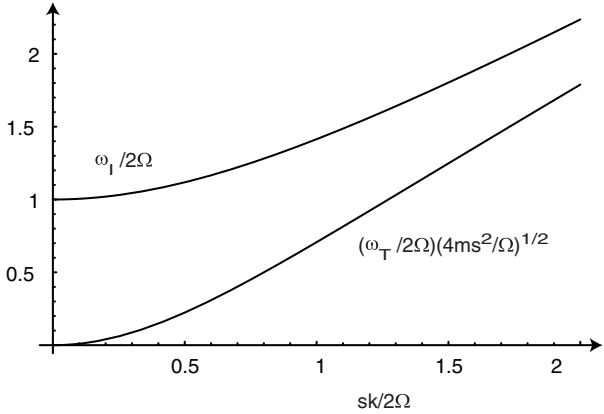,height=5.5cm}
\end{center}
\caption
{The inertial and Tkachenko mode frequencies vs. wavevector measured
in units of $2\Omega/s$.  The inertial mode (upper curve) is in units
of $2\Omega$, while the Tkachenko mode (lower curve) is in units of
$(\Omega^3/ms^2)^{1/2}$.  The modes are calculated for $C_1,C_2$ in the
Thomas-Fermi regime.  In the quantum Hall regime, the Tkachenko modes are
softer by a factor $(9/\pi^2)(ms^2/10\Omega)^{1/2}$.}
\label{FIG1}
\end{figure}

    The very soft Tkachenko mode in the rapidly rotating regime leads to
infrared singular behavior in the vortex transverse displacement-displacement
correlations at finite temperature, and in the order parameter phase
correlations even at zero temperature \cite{SHM,qhmodes}.  In a finite system
the single particle density matrix, $\langle\psi(r)\psi^\dagger(r')\rangle$,
falls algebraically as $(4N_v/\pi^2)^{-\eta}$ for $|r-r'|\sim R$, where $\eta
\simeq \pi^2\sqrt{10}N_v/9N$, $N$ is the total particle number, and $N_v$ is
the total number of vortices present \cite{qhmodes}.  However, dephasing of
the condensate only becomes significant as $N_v\to N$, and not
necessarily before the vortex lattice melts.

\begin{figure}
\begin{center}
\epsfig{file=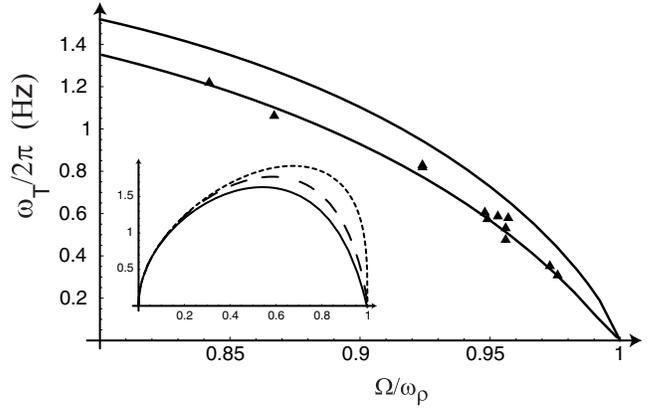,height=5.4cm}
\end{center}
\caption
{Frequency of the lowest Tkachenko mode.  The upper curve is for constant
shear modulus, $C_2$, while the lower curve includes a decreasing $C_2$,
(Eq.~(\ref{C2drop})) with $\gamma=4$.  The data (triangles), from
Ref.~[16], are multiplied by a factor $(N/2.5\times 10^6)^{-2/5}$ to
compare with theory, which is calculated for $N=2.5\times 10^6$.  The inset
shows the Tkachenko frequency over the entire range of $\Omega$; the upper
(short dashed) curve is the mode frequency to lowest order in the wavevector,
the middle (dashed) curve includes the full frequency dependence at constant
$C_2$, and the lower curve includes the decreasing $C_2$ as well.}
\label{FIG2}
\end{figure}

    To compare Eq.~(\ref{tk}) with the experiment of Ref.~\cite{jilatk}, we
extract the effective value of $k$ in the lowest Tkachenko mode from the
numerical result of Ref.  \cite{crescimanno} for the fundamental frequency:
$\omega_T \simeq 1.43a\Omega/R = 2.72(\Omega/m)^{1/2}/R$, where
$a=(2\pi)^{1/2}/3^{1/4}(m\Omega)^{1/2}$ is the lattice constant, and $R$ the
Thomas-Fermi radius in the transverse direction.  Comparison with the slow
rotation Tkachenko frequency, $(\Omega/4m)^{1/2}k$ indicates that the
effective wavevector of the lowest mode is $k_0=\alpha/R$, where $\alpha
\simeq 5.45$, the value of $k$ we use in the following.  The transverse
Thomas-Fermi radius of a rotating BEC \cite{FG} is
\beq
  R^2= d^2\tau (1-x)^{-3/5},
\eeq
where $x=\Omega^2/\omega_\rho^2$, $d=1/\sqrt{m\omega_\rho}$ is the
transverse oscillator length, $\tau =
[(15Nba_s/d)(\omega_z/\omega_\rho)]^{2/5}$, and $\omega_z$ is the axial
trapping frequency.  The sound velocity, evaluated in the center of the trap
(for simplicity), is
\beq
 ms^2 = gbn(0) = \frac{\omega_\rho}{2}\tau(1-x)^{2/5},
\eeq
so that
\beq
  \frac{\Omega}{ms^2}= \frac{2}{\tau}\frac{x^{1/2}}{(1-x)^{2/5}}.
\eeq
Also, $(\Omega/sk_0)^2 = 2x/(1-x)\alpha^2$; in the present
experiments, $\Omega/sk_0$ reaches $\sim 1.15$.  Ignoring $b\simeq 1$ here
\cite{sound}, as well as the small $C_1+C_2$ term in Eq.~(\ref{tk}), we have,
\beq
  \omega_T^2 = \omega_\rho^2 \frac{\alpha^2}{4\tau}\,
       \frac{x^{1/2}(1-x)^{8/5}}{1-x(1-8/\alpha^2)}.
\label{tkmodes}
\eeq

    The inset in Fig. 2 shows the Tkachenko mode frequencies as a function of
$\Omega/\omega_\rho$, illustrating the initial square root rise and the
eventual falloff for $\Omega\simle\omega_\rho$.  The curves are evaluated for
the parameters of Ref.~\cite{jilatk}, $(\omega_\rho,\omega_z) = 2\pi(8.3,5.2)$
Hz, at the representative condensate number, $N=2.5\times 10^6$.  Figure 2
proper shows these curves in the region measured in \cite{jilatk}.  The upper
curve is calculated from Eq.~(\ref{tkmodes}) while the lower includes the
decrease (\ref{C2drop}) in the shear modulus of the vortex lattice, with
$\gamma=4$.  The experiments of \cite{jilatk} are at particle numbers $\sim
(0.7-3)\times 10^7$.  According to Eq.~(\ref{tkmodes}), the frequencies scale
as $N^{-2/5}$.  Thus to facilitate comparison with theory, shown for
$N=2.5\times 10^6$, we have scaled the individual data points down by a factor
$(N/2.5\times 10^6)^{2/5}$, which is equivalent to scaling the theory up by
the same factor.  The JILA experiments provide evidence for the frequency
dependence of the shear modulus of the vortex lattice.

    I am grateful to Eric Cornell, Ian Coddington, and Volker Schweikhard for
helpful correspondence as well as giving me permission to include the JILA
data here.  I also thank Sandro Stringari, Marco Cozzini, and Drew Gifford for
useful discussions.  This work was supported in part by NSF Grant PHY00-98353.

\end{multicols}

\begin{references}
\bibitem{curl} Although the local superfluid velocity in a rotating fluid
is irrotational, its long wavelength average in not\cite{BC}.
\bibitem{sound} In a weakly interacting homogeneous gas, $s^2 = gn/m$;
however here one must take into account the correction to the interaction
energy arising from the non-uniformity of the density around each vortex.  As
derived in \cite{FG}, the energy density is renormalized to $gn^2b/2$, where
$n$ is the mean density in a unit cell of the lattice, and $b=\langle
n^2\rangle/n^2$ is a numerical correction $\simge 1$, dependent on the details
of the density distribution in the cell.  At slow rotation, $b=1$, while in
the quantum Hall regime $b=1.16$ numerically,\cite{SHM}.  Neglect of the
dependence of $b$ on the mean density implies $ms^2 = gnb$, the result we use
here.
\bibitem{3d} In three dimensions, the number of degrees of freedom remains
four, since Eq.~(\ref{vphase}) now implies two constraints on the six degrees
of freedom, $n$, $v$, and $\epsilon$, and system still supports only two long
wavelength modes.  As the wavevector becomes more aligned with the rotation
axis, the Tkachenko mode tends to stiffen while the inertial mode tends to
soften.  In both modes the vortex motion remains elliptically polarized with
the longitudinal component $\simeq -i\omega/2\Omega$ times the transverse
component \cite{drew}.
\end{references}
\end{document}